\documentclass[a4paper,pdftex,floatfix,groupedaddress,pra,showpacs,twocolumn]{revtex4}
\pdfoutput=1
\usepackage{amsfonts,amsmath,amssymb}
\usepackage{graphicx}
\usepackage{bm}
\usepackage{mathrsfs}
\usepackage{textcomp}

\newcommand{\fref}[1]{Fig.~\ref{#1}}
\newcommand{\Fref}[1]{Figure~\ref{#1}}

\newlength{\figwidth}
\newlength{\figwidthsmall}
\begin{document}

\title{Ionization of oriented carbonyl-sulfide molecules by intense circularly polarized laser pulses}

\author{Darko Dimitrovski}
\author{Mahmoud Abu-samha}
\author{Lars Bojer Madsen}
\email[Corresponding author: ]{bojer@phys.au.dk}
\affiliation{Lundbeck Foundation Theoretical Center for Quantum
System Research, Department of Physics and Astronomy, Aarhus University, DK-8000 Aarhus C, Denmark}

\author{Frank Filsinger}%
\author{Gerard Meijer}%
\author{Jochen K\"upper}%
\email[Corresponding author: ]{jochen@fhi-berlin.mpg.de}%
\affiliation{Fritz-Haber-Institut der Max-Planck-Gesellschaft, Faradayweg 4-6, D-14195 Berlin,
   Germany}

\author{Lotte Holmegaard$^1$}%
\author{Line Kalh{\o}j$^1$}%
\author{Jens H. Nielsen$^2$}%
\author{Henrik Stapelfeldt$^{1,3}$}%
\email[Corresponding author: ]{henriks@chem.au.dk}%
\affiliation{$^1$\,Department of Chemistry, Aarhus University, DK-8000 Aarhus C, Denmark \\
$^2$\,Department of Physics, Aarhus University, 8000 Aarhus C, Denmark \\
   $^3$\,Interdisciplinary Nanoscience Center (iNANO),  Aarhus University, 8000 Aarhus C,
   Denmark}

\date{\today}

\begin{abstract}
We present combined experimental and theoretical results on strong-field ionization of oriented carbonyl-sulphide molecules by circularly-polarized laser pulses. The obtained molecular frame photoelectron angular distributions show pronounced asymmetries perpendicular to the direction of the molecular electric dipole moment. These findings are explained by a tunneling model invoking the laser-induced Stark shifts associated with the dipoles and polarizabilities of the molecule and its unrelaxed cation. The focus of the present article is to understand the strong-field ionization of one-dimensionally-oriented polar molecules, in particular asymmetries in the emission direction of the photoelectrons. In the following article (Phys.~Rev.~A \textbf{83}, 023406 (2011)) the focus is to understand strong-field ionization from three-dimensionally-oriented asymmetric top molecules, in particular the suppression of electron emission in nodal planes of molecular orbitals.
\end{abstract}

\maketitle


\clearpage
\section{Introduction}

When molecules are exposed to intense femtosecond laser pulses they ionize. If the ionizing laser pulse is linearly polarized the electron can be steered back to rescatter on the ion left behind, thereby initiating phenomena such as high harmonic generation, above threshold ionization, and double ionization (see, e.g., Refs.\cite{Corkum:1993:PRL,brabec:2000:rmp,Lein:jpb:2007,krausz_rmp_2009}). Being the event that initiates these central strong-field processes, ionization has attracted special attention and much effort has gone into describing and understanding it. Because molecules are not spherically
symmetric
   the ionization probability and the emission direction of the electron depends on the relative orientation between the molecule and the polarization vector of the laser pulse \cite{becker:jpb:2003,kjeldsen:2003:pra,litvinyuk:2003,kjeldsen:2004:jpb,kjeldsen:2005:pra,pinkham:2005:pra,pavicic:2007:prl,kampta2007,Telnov2007,Kumarappan:PRL:2008,Staudte:prl:2009,fernandez2009III,Spanner:pra:2009,petretti:2010:prl}. Knowledge of this orientational dependence is important for understanding, optimizing, or utilizing subsequent strong-field processes \cite{Lein:jpb:2007,krausz_rmp_2009}.

The ability to align molecules, i.e., to confine one or more molecular axes along space-fixed axes, has over the past few years provided a valuable tool to experimentally explore the orientational dependence of strong-field ionization, and consequently, opened up for comparing theoretical and experimental results \cite{litvinyuk:2003,pinkham:2005:pra,pavicic:2007:prl,Kumarappan:PRL:2008,Staudte:prl:2009,Abu-samhaPRA09}. The majority of studies have focused on nonpolar linear molecules where nonadiabatic alignment, by a linearly polarized laser pulse, provides a convenient way to prepare a sample of 1-dimensionally aligned molecules \cite{rosca-pruna:2001:prl,stapelfeldt:2003:rmp}. Most molecules are, however, polar, i.e., they do not exhibit inversion symmetry. The experimental investigation of such systems requires that not only the axes of the molecule are confined, the permanent electric dipole moment must also point in a particular direction. Thus, the molecule should be oriented in addition to being aligned. Orientation can be achieved by static electric field methods such as hexapole focusing \cite{parkerbernstein:1989:arpc,janssen:1997:jpc} and brute-force orientation \cite{loesch:1990:jcp}, by optical methods based on two-color laser fields \cite{De:prl:2009,Oda:PRL:2010}, or by combined laser and static electric field methods \cite{Goban:PRL:2008,ghafur:natphys:2009,Holmegaard:PRL102:023001,Filsinger:JCP:2009}. In the present work we employ the method relying on mixed laser and static electric fields since it provides very high degrees of alignment and orientation.

In detail we present a combined experimental and theoretical study of single ionization of a polar linear molecule, carbonyl-sulfide (OCS), by near infrared 30 femtosecond laser pulses. First results were recently presented elsewhere \cite{Holmegaard:NatPhys:2010}. Unlike studies aimed at recollision phenomena, e.g., high-order-harmonic generation, circularly polarized pulses are employed. By doing so the strong-field dynamics is simplified since the circularly polarized field drives the electrons away from the parent molecule and thus turns off recollision. Our studies focus on the photoelectron angular distributions (PADs) from single ionization. When the OCS molecules are tightly aligned and oriented pronounced asymmetries are observed in the experimental PADs perpendicular to the fixed molecular axis. The asymmetries are absent for randomly oriented molecules. Our theoretical analysis, based on a modified tunneling theory, rationalizes the experimental findings and shows that the observed asymmetries are determined by the difference in ionization probability between the cases when the circularly polarized field points in the same and in the opposite direction as the permanent dipole moment \cite{CEP-note}. Notably, the PADs reflect the permanent dipole moment and the polarizability of the active molecular orbital, which again, in the case of the highest-occupied molecular orbital (HOMO), may be related to the dipole moments and polarizabilities of the neutral molecule as well as of its unrelaxed cation. The calculated results are exponentially sensitive to (temporal) changes in these quantities and hence point to the extension to time-resolved measurements of valence electron dynamics using pump-probe settings.

The paper is organized as follows: In Sec.~II the experimental technique is discussed and in Sec.~III, the experimental results are presented. Section IV presents the theory and in Sec.~V the theoretical predictions are compared to the experiment. Conclusions are given in the last section. Appendix A provides a summary of the molecular properties of OCS and its cation relavant for the present study.

\section{Experimental method}

The experimental setup is described in detail elsewhere \cite{BN_circ} so the discussion here is brief. A gas mixture of~$\sim$10~mbar carbonylsulfide (OCS) and 10 bar
of Ne is expanded supersonically into vacuum through an Even-Lavie valve \cite{even:2000:jcp,even:2003:jcp} forming a pulsed molecular beam. The molecular beam is skimmed twice before entering a 15-cm-long electrostatic deflector that spatially disperses the molecular beam in the vertical direction according to the quantum states populated \cite{Holmegaard:PRL102:023001,Filsinger:JCP:2009}. After exiting the deflector the molecular beam is crossed at 90$^\circ$ by two focused laser beams, one to align and orient the molecules and one to induce ionization. The experiments described here are conducted on the most deflected molecules, i.e., a subset of molecules
selected in the lowest lying rotational quantum states \cite{Holmegaard:PRL102:023001,Filsinger:JCP:2009}.

The alignment beam originates from an injection seeded Q-switched Nd:YAG laser (20 Hz, $\tau_\text{FWHM}~=~10~\text{ns}$, $\lambda~=~1064~\text{nm}$). The spotsize, ($\omega_0 ^{\text{YAG}}$, in the focus at the crossing with the molecular beam is 34~$\mu$m, yielding a peak intensity of $\sim$~8$\times10^{11}$~W/cm$^2$. The ionization laser beam, termed the probe beam, originates from a pulsed femtosecond Ti-Sapphire system (1 kHz, $\lambda~=~800~\text{nm}$) externally compressed to 30 fs (FWHM) and focused to $\omega_0 ^{\text{probe}} = 21~\mu$m resulting in a peak intensity of $\sim$~5.4$\times10^{14}$~W/cm$^2$. The probe pulse is electronically synchronized to the peak of the YAG pulse, where the degree of alignment is highest.

The ions or electrons produced by the probe pulses are extracted with a weak static electric field in a velocity map imaging (VMI) geometry and projected onto a two dimensional detector consisting of a micro channel plate (MCP) detector backed by a phosphor screen. The ion or electron images on the
phosphor screen are recorded by a CCD camera and the coordinates of each individual particle hit are determined.

\section{Experimental results}

\subsection{Alignment and orientation}
\label{alignment-orientation}

\begin{figure}
    \centering
   \includegraphics[width=\figwidthsmall]{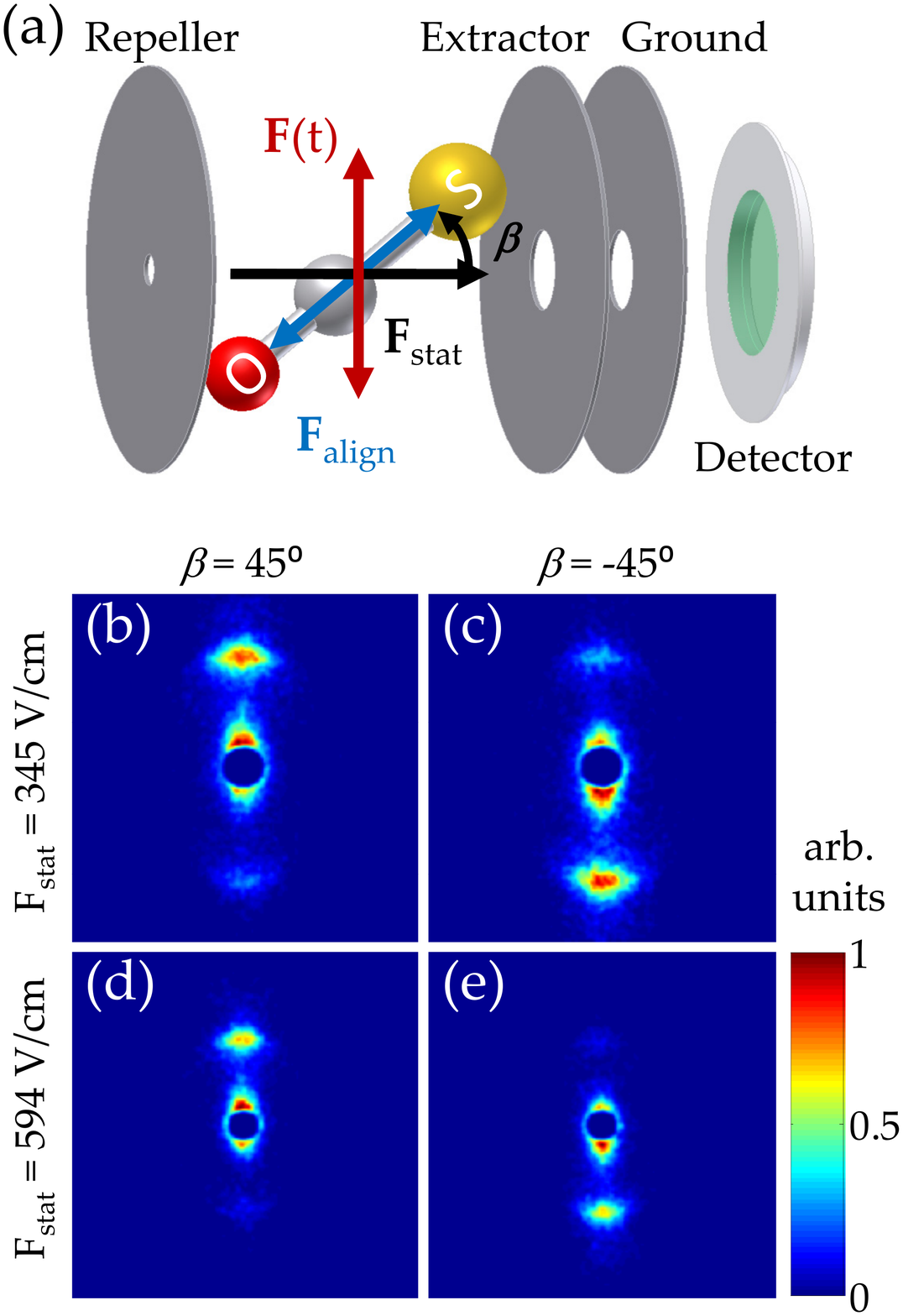}
   \caption{(Color online) (a) Schematic of the velocity map imaging spectrometer used
 to detect ions or electrons. The alignment of the molecules, illustrated by the OCS model,
 is determined by the alignment laser polarization, here shown for  $\beta~=~45^\circ$.
 The static electric field of the spectrometer, pointing from the repeller to the extractor
 electrode for ion detection, breaks the head-for-tail symmetry by preferentially placing
 the O-end towards the repeller. When detecting electrons
 the polarity of the electrodes is inverted forcing the S-end of the molecules
 towards the repeller electrode. (b), (c) Images of S$^+$ ions for $\beta~=~45^\circ$ and -45$^\circ$, respectively and $\text{F}_{\text{stat}}~=~345$ V/cm. In (b) 73~\% of
 all S$^+$ ions appear in the upper half of the detector. In (c) 28~\% of all S$^+$ ions appear
  in the upper half of the detector.  (d), (e) Images of S$^+$ ions for $\beta~=~45^\circ$ and -45$^\circ$, respectively with the static field increased to $\text{F}_{\text{stat}}~=~594$ V/cm. In (d) and (e) respectively 80~\% and 19~\% of
 all S$^+$ ions appear in the upper part of the detector. The intensity of the probe laser is $\sim$~5.4$\times10^{14}$~W/cm$^2$.\\
 }
  \label{fig:setup}
\end{figure}

The target of adiabatically oriented molecules is obtained by the combined action of the ac electric field from the YAG pulse and the weak static electric field present in the VMI spectrometer as shown in previous studies \cite{Holmegaard:PRL102:023001,Filsinger:JCP:2009}. The strongest orientation is obtained when the YAG pulse is polarized along static electric field, $\textbf{F}_{\text{stat}}$ and this is the geometry used in the photoelectron studies presented in Sec. \ref{sec:exp:PAD}. To characterize orientation using 2D ion imaging, as we do here, it is, however, necessary to rotate the molecules away from $\textbf{F}_{\text{stat}}$ as shown in Fig. \ref{fig:setup}(a). Our method to characterize alignment and orientation is based on Coulomb exploding the molecules with an intense probe pulse and subsequently recording the velocities of the recoiling ions by the 2D imaging detector. This method does not work well for molecules aligned along $\textbf{F}_{\text{stat}}$ because all recoiling ions tend to collapse in the center of the detector. Therefore, the measurements of alignment and orientation are performed for molecules with their molecular axis rotated 45 degrees away from $\textbf{F}_{\text{stat}}$.

\Fref{fig:setup}(b) and (c) display S$^+$ ion images from Coulomb explosion of the OCS molecules with the probe pulse
linearly polarized vertically and the YAG pulse polarized at $\beta~=~45^\circ$ or -45$^\circ$, where
 $\beta$ is the angle between $\textbf{F}_{\text{stat}}$ and the alignment laser
field, $\textbf{F}_{\text{align}}$ [see Fig. \ref{fig:setup} (a)]. The amplitude of the static field is 345~V/cm. We interpret the S$^+$ ions detected at small radii, near the center of the images, as originating from OCS molecules, singly ionized by the probe pulse and dissociating into CO and S$^+$. By contrast, the S$^+$ ions in the pair of radially and angularly localized regions (at the outermost part of the images) is interpreted as originating from OCS molecules, doubly ionized by the probe pulse and subsequently fragmenting into a CO$^+$-S$^+$ ion pair. The recoil of S$^+$ ions from this Coulomb explosion channel reflects the direction of OCS at the moment of ionization and is thus a useful experimental observable to determine the molecular alignment and orientation.

The strong angular confinement of the S$^+$ from the Coulomb explosion channel shows that the OCS molecules are sharply 1-dimensionally aligned along the
polarization of the linearly polarized YAG pulse. In addition, a pronounced asymmetry of the S$^{+}$
ions emitted either parallel or anti-parallel to $\textbf{F}_{\text{stat}}$, with an excess of S$^+$ in the
upper (lower) region for $\beta~=~45^\circ~(-45^\circ)$, shows that the molecules are oriented with the S-end
preferentially pointing toward the extractor electrode where the electrical potential is lowest. These findings
are fully consistent with recent alignment and orientation studies on iodobenzene \cite{Holmegaard:PRL102:023001,Filsinger:JCP:2009}
and 2,6-difluoroiodobenzene \cite{Nevo:PCCP:2009} as well as with former mixed-field orientation studies on OCS \cite{minemoto:2003:jcp}. In Figs. \ref{fig:setup}(d) and (e) the static electric field is increased to $\text{F}_{\text{stat}}~=~594$~V/cm resulting in a clear improvement of the degree of orientation.

The theoretical treatment of the experimental PAD measurements, presented below,
requires knowledge of the degree of orientation, i.e., the fraction of molecules with the
S-end pointing towards the repeller. The ion imaging measurements occur for beta = 45 degrees rather than the 0 degree geometry used in the PAD measurements. To, nevertheless, provide an estimate of the degree of orientation we note that E$_{\text{stat}}$ = 345~V/cm in the PAD measurements [See Sec. \ref{sec:exp:PAD}].
This value falls in between the value of the effective static field, i.e., E$_{\text{stat}}$ along the
 OCS bond axis, of Figs. 1(b) and 1(c) ($\sim$~cos(45$^\circ$) $\times$~345~V/cm = 244~V/cm) and Figs. 1(d) and
1(e) ($\sim$~cos(45$^\circ$) $\times$~595~V/cm = 420~V/cm). In the former (latter) case the orientation corresponds
to a 73\% (80\%) up-to-total ratio. Therefore, the orientation in the PAD experiment geometry should
be at least 77-78 \%. The vertical probe geometry applied in Fig. \ref{fig:setup} does, however, underestimate the degree
of orientation, because the probe pulse preferentially ionizes (probes) the molecules aligned along its vertical
polarization axis where the static field goes to zero and the molecules are, therefore, only weakly oriented. As a consequence, we estimate
that the orientation in the PAD geometry corresponds to 80 \% of the molecules having their O-end toward the detector, see Fig. \ref{fig:PAD:schematic}.

\subsection{PADs from single ionization of OCS}
\label{sec:exp:PAD}

\begin{figure}
    \centering
   \includegraphics[width=\figwidthsmall]{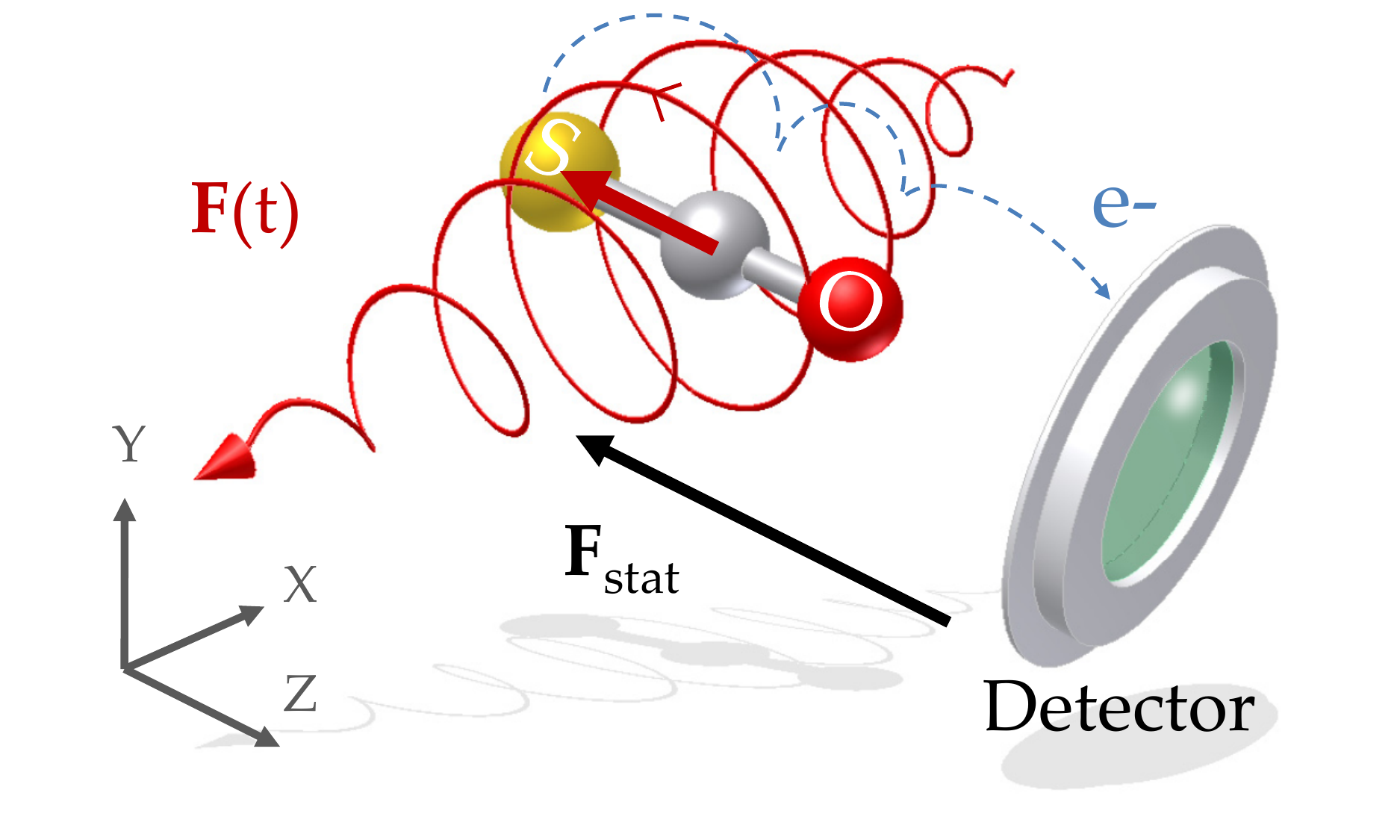}
 \caption{(Color online) Schematic of the experimental setup showing an OCS molecule oriented with its permanent dipole moment (bold red arrow) pointing in the direction of the static electric field. The left circularly-polarized (LCP) probe pulse ionizes the molecule and imparts an upward momentum to the freed electron resulting in recording on the upper part of the detector (see text for details). }
  \label{fig:PAD:schematic}
\end{figure}

For the PAD experiments the same experimental setup, described in
Sec.\ref{alignment-orientation} is used, but some essential parameters are changed. The polarization
state of the 30 fs probe pulses, denoted as $\mathbf{F} (t)$ is changed from linear to circular and the intensity is lowered to
$\simeq2.4\times10^{14}$~W/cm$^2$ corresponding to a regime where the OCS molecules only undergo single
ionization with essentially no fragmentation. The intensity puts the dynamics in the tunneling regime~\cite{Keldysh} and the circular polarization ensures that no recollision of the freed electron with its parent ion occurs. Both conditions are important for the interpretation and modeling of the observed PADs. Also, the polarization of the alignment pulse is changed such that its major axis is parallel to the static field axis. Furthermore, to extract
electrons instead of ions in the PAD measurements the polarity of the velocity map imaging
spectrometer is inverted. Hereby, the OCS molecules are confined along the static field axis with the
O-end facing the detector (See \fref{fig:PAD:schematic}).

\begin{figure}
    \centering
   \includegraphics[width = \figwidth]{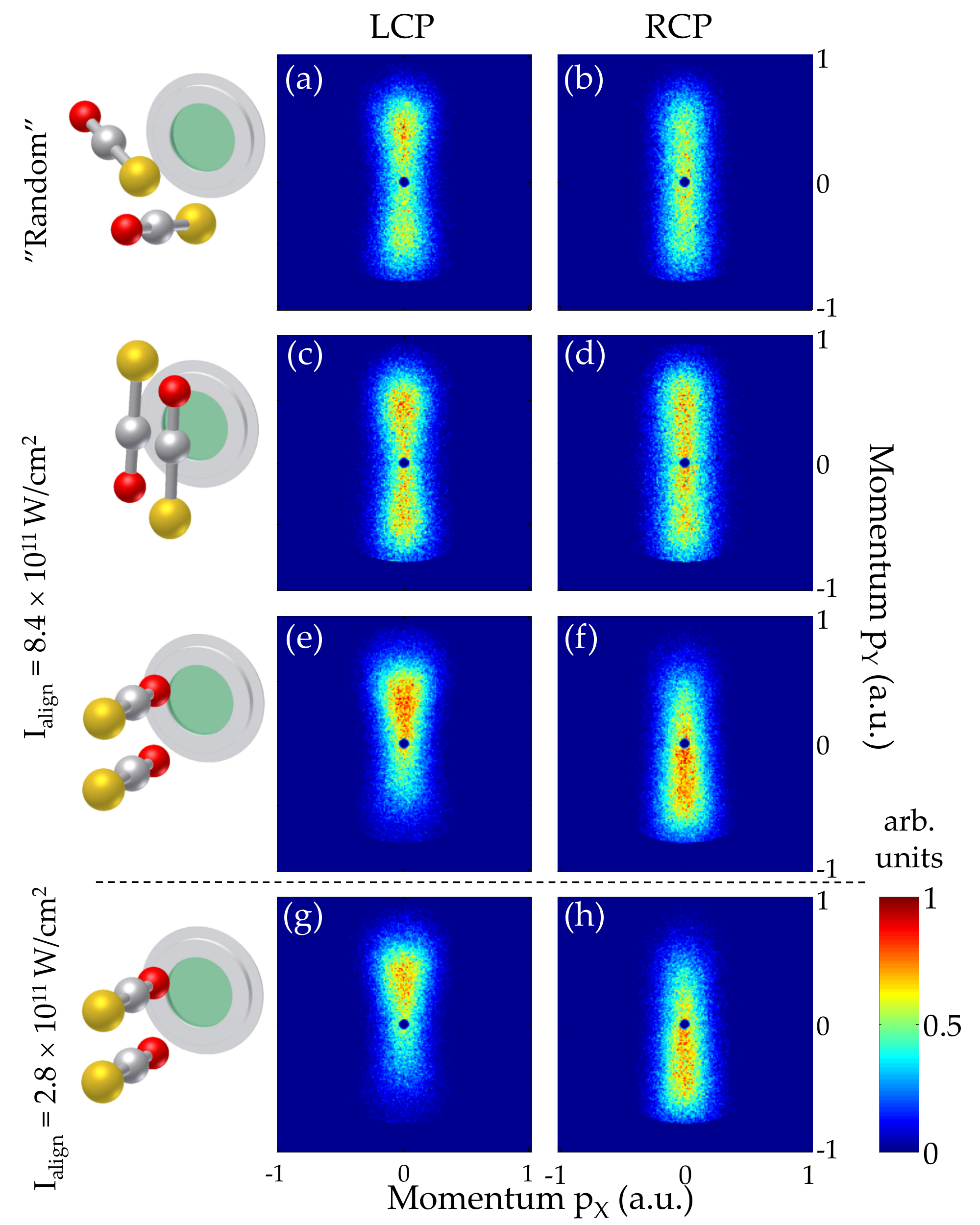}
 \caption{
  (Color online) Two-dimensional momentum image of electrons produced when a randomly oriented sample of OCS molecules are ionized by (a) the LCP  probe pulse. The polarization plane of the probe pulse is in the (Y,Z) plane, i.e. perpendicular to the image (detector). (b) Same as (a) but for a RCP probe pulse. (c) and (d) as (a) and (b) but with the OCS molecules aligned along the $Y$-direction by the alignment pulse polarized parallel to the image plane. (e) and (f) as (a) and (b) but with the OCS molecules aligned along the $Z$-direction by the alignment pulse polarized perpendicular to the image plane. (g) and (h) as (e) and (f) but with the intensity of the alignment pulse lowered from $\text{I}_{\text{align, YAG}}~=~8.4\times10^{11}$~W/cm$^2$ to $\text{I}_{\text{align, YAG}}~=~2.8\times10^{11}$~W/cm$^2$. The intensity of the 800 nm 30 fs probe pulse is kept at $2.44 \times 10^{14}$ W/cm$^2$ in all pictures.}
  \label{fig:OCS:PADs}
\end{figure}

The electron images are shown in \fref{fig:OCS:PADs}. With only the probe pulse [Figs. \ref{fig:OCS:PADs}(a), \ref{fig:OCS:PADs}(b)] the electrons emerge in a stripe parallel to the ($Y$,$Z$) polarization plane of the $\simeq 2.4 \times 10^{14}$ W/cm$^2$ probe pulse. The images are essentially up-down symmetric and the marginal difference between the images obtained with left and right circularly  polarized (LCP and RCP) pulses is due to experimental imperfections in the purity of the polarization state and a weak orientation of the molecules caused by the static field alone \cite{Nevo:PCCP:2009}. When the molecules are one-dimensionally (1D) aligned along the $Y$-direction, i.e., the molecular axis is confined along the $Y$-axis but with no preferred direction of the dipole moment, not much happens and no up-down asymmetry is observed [Figs. \ref{fig:OCS:PADs}(c) and \ref{fig:OCS:PADs}(d)]. When the YAG pulse polarization is turned parallel to $\textbf{F}_{\text{stat}}$, and the molecules thus become 1D aligned and oriented, a strong up-down asymmetry is observed  [Figs. \ref{fig:OCS:PADs}(e), \ref{fig:OCS:PADs}(f)]. The asymmetry reverses as the helicity of the probe pulses is flipped. For LCP (RCP) probe pulses the number of electrons detected in the upper part compared to the total number in the image is $\sim$64\%  (39\%).

To investigate if the YAG pulse not only induces molecular alignment and orientation but also influences the photoelectron trajectories PADs were measured under experimental conditions identical to those used in Figs. \ref{fig:OCS:PADs} (e) and \ref{fig:OCS:PADs} (f) but with the YAG intensity reduced by a factor of three. Independent measurements, using $\text{S}^+$ ion imaging (not shown here) showed that the degrees of alignment and orientation remain almost unchanged. The resulting electron images are shown in Figs. \ref{fig:OCS:PADs} (g) and \ref{fig:OCS:PADs} (h). They are very similar to those obtained with the images obtained at three times higher YAG pulse intensity. In particular the up-to-total number of electrons is $\sim$66\%  (38\%) in Figs. \ref{fig:OCS:PADs} (g) and \ref{fig:OCS:PADs} (h), respectively, which is almost the same as for Figs. \ref{fig:OCS:PADs} (e) and \ref{fig:OCS:PADs} (f), strongly indicating that the YAG pulse does not cause any significant distortion of the electron trajectories. This is corroborated by measurement on benzonitrile where experiments were conducted on molecules at higher rotational temperatures and without state selection \cite{BN_circ}. In that case no up-down asymmetry of the photoelectrons is observed even at the highest YAG pulse intensity. We conclude that the YAG pulse together with the static electric field serve to control the alignment and orientation of the molecules but does not otherwise visibly influence the ionization process by the probe or the subsequent trajectories of the released electrons.

\begin{figure}
    \centering
   \includegraphics[width = \figwidth]{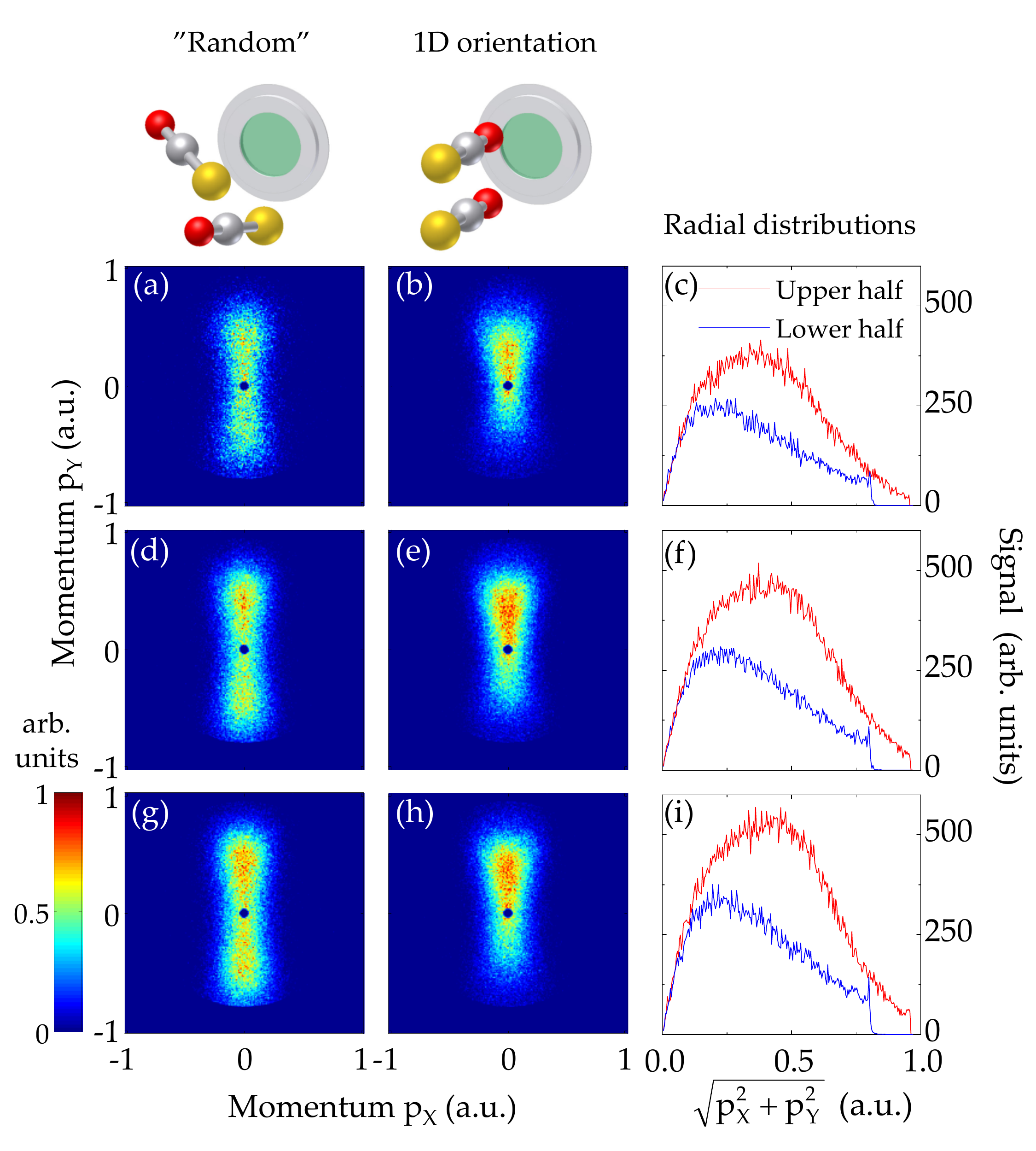}
 \caption{
 (Color online) Two-dimensional momentum image of electrons produced when a randomly oriented sample of OCS molecules is ionized by the LCP for increasing intensity of the ionization pulse, corresponding to $\text{I}_{\text{probe}}~=~1.76\times10^{14}$~W/cm$^2$ in (a), $\text{I}_{\text{probe}}~=~2.44\times10^{14}$~W/cm$^2$ in (d) and $\text{I}_{\text{probe}}~=~2.83\times10^{14}$~W/cm$^2$ in (g). (b), (e) and (h), same as (a), (d) and (g), respectively, but with the OCS molecules aligned along the $Z$-direction by the alignment pulse polarized perpendicular to the image plane. In (c), (f) and (i) the radial distributions of the upper and lower parts of images (b), (e) and (h) are shown. $\text{I}_{\text{align, YAG}}~=~8.4\times10^{11}$~W/cm$^2$\\}
  \label{fig:OCS:PADs:intensities}
\end{figure}

To investigate the role of the intensity of the probe laser pulse, measurements at three different intensities were performed. \Fref{fig:OCS:PADs:intensities} shows the electron images obtained with a left circularly polarized pulse. In Figs. \ref{fig:OCS:PADs:intensities}(a), \ref{fig:OCS:PADs:intensities}(d) and \ref{fig:OCS:PADs:intensities}(g) only the probe pulse is included for increasing intensity corresponding to $\text{I}_{\text{probe}}~=~1.76\times10^{14}$~W/cm$^2$ in (a), $2.44\times10^{14}$~W/cm$^2$ in (d) and $2.83\times10^{14}$~W/cm$^2$ in (g). In Figs. \ref{fig:OCS:PADs:intensities}(b), \ref{fig:OCS:PADs:intensities}(e) and \ref{fig:OCS:PADs:intensities}(h) the molecules are 1D aligned and oriented and again clear up/down asymmetries are observed in the photoelectron distributions. In Figs. \ref{fig:OCS:PADs:intensities}(c), \ref{fig:OCS:PADs:intensities}(f) and \ref{fig:OCS:PADs:intensities}(i) the corresponding radial distributions, obtained by angularly integrating the images, are given for the upper and lower half of Figs. \ref{fig:OCS:PADs:intensities}(b), \ref{fig:OCS:PADs:intensities}(e) and \ref{fig:OCS:PADs:intensities}(h), respectively. It is seen that as the probe intensity is increased the number of electrons detected increases, and they acquire more momentum, i.e., extend towards the edge of the detector. The maximum momentum observed is limited by the size of the detector. The number of electrons appearing in the upper half of the images compared to the lower half is almost unchanged for increasing intensity with $\sim$63\% in Fig. \ref{fig:OCS:PADs:intensities}(b) and $\sim$64\% in Figs. \ref{fig:OCS:PADs:intensities}(e) and \ref{fig:OCS:PADs:intensities}(h). These experimental observations are compared to the calculated results in Sec. \ref{comparison}

\section{Theory}
\label{theory}

For comparison with theory, we focus on the intensities
$2.44 \times 10^{14}$ and $2.83 \times 10^{14}$ W/cm$^2$ for the 800 nm, 30 fs pulses.
For OCS, these laser parameters result in Keldysh parameters ~\cite{Keldysh} $\gamma = \omega \sqrt{2 I_p } /F $ of $\gamma=0.87$ and $\gamma =0.82$ that are both lower than unity, so it is justified to
use tunneling theory to describe the photoelectron emission process. The
existing tunneling models, however, need to be modified to correctly
describe ionization from a polar molecule
with large dipole moments and polarizabilities such as OCS.

In a circularly polarized field the electron is driven away from
the (unrelaxed) cation, which is in contrast to the case of a linearly
polarized field where rescattering and post-ionization interaction
are important. This fact simplifies the propagation after the
initial ionization step in circularly polarized fields: it proves
sufficient to propagate classical equations of motion for the
electron in the external field ignoring the effect of the molecular potential. The full-width at half
maximum of the 800 nm laser pulse used in the experiment is 30 fs,
and accordingly there are more than 10 cycles within the envelope.
Therefore it is sufficient to model the laser pulse of the experiment
by a periodic field with constant  amplitude, and assume that it is switched off adiabatically
in the long time limit. We focus on the case of a left circularly polarized laser
pulse (LCP), and define the electric field $\mathbf{F} (t)$ of the
probe pulse as
\begin{equation}
\mathbf{F} (t) = F_0 \sin ( \omega t) \mathbf{\hat{e}}_y + F_0 \cos ( \omega t ) \mathbf{\hat{e}}_z .
\label{eq:9}
\end{equation}
In the above equation $F_0$ is the field amplitude, $\omega$ is
the angular frequency and the product $\theta = \omega t $ is the
angle between the electric field vector and the positive $Z$-axis.
The vector potential corresponding to \eqref{eq:9} is ($\mathbf{F}(t) = - \partial_t \mathbf{A}(t))$
\begin{equation}
\mathbf{A} (t) = \frac{F_0}{\omega} \cos ( \omega t) \mathbf{\hat{e}}_y - \frac{F_0}{\omega} \sin ( \omega t ) \mathbf{\hat{e}}_z .
\label{eq:10}
\end{equation}

Assuming that there is no influence of the molecular potential on
the final momentum of the escaping continuum electron, the emission at time
$t_0$ (angle ${\theta} = \omega t_0 $) creates an electron with
final momentum in the ($Y$,$Z$) plane equal to
\begin{align}
p_Y = &- A_Y ( {\theta} / \omega ) = - \frac{F_0}{\omega} \cos ( {\theta} ) \notag \\
p_Z = &- A_Z ( {\theta}  / \omega ) =  \frac{F_0}{\omega}   \sin ( {\theta} ) .
\label{eq:12}
\end{align}

\begin{figure}
{\includegraphics[width=\figwidthsmall]{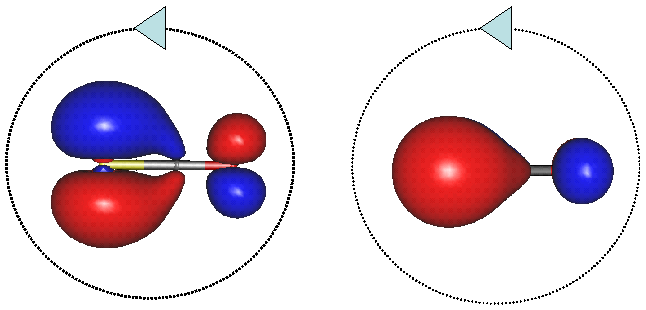}}
\caption{(Color online) Visualization of the degenerate HOMO orbitals of OCS, at
an iso-density contour value of 0.1. The S-end of the molecule is
on the left in these images. The orbital to the left lies in the
polarization plane of the laser field, and contributes more to the total ionization yield than the orbital to
the right, which has a nodal structure in the polarization plane and is obtained by 90 degrees rotation  of the orbital to the left around the molecular axis.
The circles with arrows pointing counter clockwise illustrate the
LCP field.} \label{fig:OCS_orbitals}
\end{figure}

From the above considerations it is
clear that any  asymmetry in the  electron emission predicted by tunneling theory
 for an  oriented polar
molecule directly translates into the up-down asymmetry
(positive-negative $Y$-components of the final electron momentum)
observed experimentally. Moreover, as it is seen from Eq. \eqref{eq:12}, the favoured detection of electrons in the
upper half of the detector for a LCP field translates into a favoured emission from the O-end
of the molecule when the field points in negative Z-direction.
An obvious candidate responsible for the
asymmetry in the electron emission is the asymmetry of the HOMO in
the asymptotic regions of large spatial distances (see Fig. \ref{fig:OCS_orbitals}) that enters into the tunneling model.
To address this question, we first briefly review the existing tunneling models.

The tunneling rate in a static field is governed by the
exponential $\exp ( - 2 {\kappa}^3  /(3 F)) $
\cite{Landau,Perelomov:1966:JETP}, where $F = |\mathbf{F}| $ and
$\kappa  = \sqrt{2 I_p (0) }$, where $I_p (0)$ is the field-free
ionization potential. In the case of an atom in its ground state, the
pre-exponential factor in the tunneling expression accounts for
the symmetry of the initial state
\cite{Perelomov:1966:JETP,Ammosov:1986:JETP}. An extension of the
tunneling theory, fully in line with the atomic case, was carried
out for the molecules \cite{CDLin2002}. This molecular tunneling
ionization theory takes the orientation
of the field relative to the molecular axis into account. The rate in molecular tunneling theory
\cite{CDLin2002} for static fields is
\begin{align}
w ( \mathbf{F} ) = & \frac{1}{
{\kappa  }^{(2Z/ \kappa  )-1}} \exp \left( - \frac{ 2 { \kappa  }^3 }{3 F} \right) \notag \\ 
&\times \sum_{m' }\frac{B^2 (m' )}{2^{|m' |} |m' |!}  { \left( \frac{2 {\kappa}^3 }{F} \right) }^{(2Z/ \kappa ) - |m' |-1} ,
\label{eq:4}
\end{align}
where $B(m')$ accounts for the orientation of the field with
respect to the molecular axis,
\begin{equation}
B(m' )= \sum_l C_{lm} D ^{l}_{m' ,m} ( \mathbf{ \hat{F}} ) Q(l,m).
 \label{eq:5}
\end{equation}
In the above equation, $D ^{l}_{m' ,m} ( \mathbf{ \hat{F}} )$ is
the Wigner rotation matrix element (see, e.g., \cite{zare:angularmomentum}), for
passive rotation of the coordinate system through angles $\mathbf{\hat{F}}$ from the molecule-fixed frame to the laboratory-fixed frame with the $Z$-axis determined by the direction of the external field. The coefficient $Q(l,m)$, given by
\begin{equation}
Q(l,m)=(-1)^{(|m|-m)/2} \sqrt{\frac{2l+1}{2} \frac{(l+|m|)!}{(l-|m|)!} } ,
\label{eq:qlm}
\end{equation}
is related to the dominant
behavior of spherical harmonics along the field direction \cite{Varshalovich},
\begin{equation}
Y_{lm} (\theta , \phi ) \approx Q(l,m) \frac{{\sin}^{|m|}( \theta )
}{2^{|m|} |m|!} \frac{\exp(im\phi)}{\sqrt{2 \pi }} . \label{eq:6}
\end{equation}
In the above equation, $\theta$ and $\phi$ are angular coordinates in the spherical
coordinate system where the $Z$-axis is directed along the field. Finally, the $C_{l,m} $ coefficients are related to the asymptotic behaviour of the wavefunction of the HOMO, that is
\begin{equation}
\Psi ( \mathbf{r} ) \approx r^{\frac{Z}{\kappa} - 1} \exp (- \kappa r ) \sum_{l,m} C_{lm} Y_{lm} ( \mathbf{\hat{r}} ).
\label{eq:asymptotic}
\end{equation}
In Eq.~\eqref{eq:asymptotic}, the radial part solves the radial Schr\"{o}dinger equation for the electron in the Coulomb field to first order in $1/r$.
The $C_{lm}$ coefficients for both degenerate HOMO orbitals  of OCS were calculated by projecting the orbitals obtained using standard quantum chemistry calculation \cite{GAMESS} onto the asymptotic form \eqref{eq:asymptotic}, yielding coefficients with $m= \pm 1$ and with $l$ up to $l=5$. Since the oriented OCS molecule is free to rotate around its molecular axis, a combined response from both orthogonal degenerate HOMO orbitals is required. The two orbitals are rotated by 90 degrees from each other along the molecular axis. We assume one is in the polarization plane of the laser pulse and one is perpendicular to it. To calculate the combined response from these two orbitals the associated angle-dependent tunneling rates of both orbitals are added incoherently since the molecule is in a mixed state with respect to the degenerate orbitals.

Direct application of the molecular tunneling theory sketched above, however, gives an \textit{opposite}
emission dependence to the one required to describe the asymmetry
observed in the experiment. Namely, the $B$ coefficients from
\eqref{eq:5} are such that the emission from the S-end is favoured
with respect to the emission from the O-end. This reflects the
simple fact that the HOMO orbitals of the OCS are such that the
wavefunction is predominantly located toward the S-end of the
molecule (see Fig. \ref{fig:OCS_orbitals}). As we
will show below, however, to describe the tunneling ionization process from  a polar molecule it is essential to take into account the angle-dependent shifts of the ionization potential, induced by the polar system. Unlike the $B$-coefficients
that depend on the geometry of HOMO and influence the
pre-exponential factor in the tunneling rate of Eq. \eqref{eq:4},
the Stark shifts affect both the pre-exponential factor and
the argument of the exponential in the tunneling rate and decide
from which end of the molecule the preferred emission occurs.

Our modification of the tunneling theory is building on the fact that in electric fields, due to the molecule's polarity and its polarizability, hyperpolarizability, etc., the energy levels of the molecule shift. These shifts are negligible for small nonpolar molecules and atoms. In the case of molecules with large dipole
moments and polarizabilities the influence of these Stark shifts
cannot be neglected and in the present quasistatic  limit they must be
included. The unrelaxed cation of the molecule in question is even
more tightly bound than the neutral molecule and the characteristic timescale for the electronic motion is shorter. Hence, if for the
molecule the field can be regarded as \textit{static}, then for
the cation the field is static as well. The total energy of a molecule (M) and its unrelaxed cation (I) $E^{M/I} ( \mathbf{F} )$ in a static field $\mathbf{F}$, up to second order in field strength is given by (see, e.g. \cite{atkins})
\begin{equation}
E^{M/I} (\mathbf{F})=E^{M/I} (0)-{\boldsymbol \mu}^{M/I} \cdot \mathbf{F} -\frac{1}{2} \mathbf{F}^\text{T} {\boldsymbol \alpha}^{M/I} \mathbf{F} ,
\label{eq:1}
\end{equation}
where ${\boldsymbol \mu}^{M/I}$ is the dipole moment, ${\boldsymbol
\alpha}^{M/I}$ is the polarizability tensor, $E^{M/I} (0)$ is the field-free
total energy of the system. The next term in the expansion of the total energy as a function of field strength involves the hyperpolarizability. In OCS, the contribution of the hyperpolarizability is negligible at the intensities used in the present experiment, so the Stark shift is due to the dipole moment and the polarizability only. The Stark shift due to the polarizability is larger than the Stark shift due to the permanent dipole moment. What is important, however, is the \textit{difference} between the total energy of the molecule and the positive ion, i.e., the ionization potential. Since the molecule and the ion do not have identical permanent dipole moments and polarizabilities, the ionization potential $I_p  = E^{I} - E^{M} $ becomes
\begin{equation}
I_p (\mathbf{F}) = I_p (0) +{\boldsymbol \Delta \mu} \cdot
\mathbf{F} +\frac{1}{2} \mathbf{F}^\text{T} {\boldsymbol \Delta} {\boldsymbol \alpha}
{\mathbf{F}} , \label{eq:2}
\end{equation}
where $ \mathbf{F}^\text{T}$ is the transpose of the field vector and
\begin{equation}
{\boldsymbol \Delta} {\boldsymbol \mu} =
{\boldsymbol \mu}^{M} - {\boldsymbol \mu}^{I}  \quad
{\boldsymbol \Delta} {\boldsymbol \alpha} = {\boldsymbol
\alpha}^{M} - {\boldsymbol \alpha}^{I} . \label{eq:3}
\end{equation}
In the above formulation, we assumed that the electron in highest occupied molecular orbital  was promoted to the continuum. In this case
the change of the ionization potential as a function of field strength can be referred as to as the Stark shift of the HOMO orbital, with the corresponding dipole moment and polarizability. Such modifications of the ionization potential of orbitals other than HOMO can be calculated as well. Note that here we do not make any distinction between the dipole moments and polarizabilities of (possibly) different ionic products obtained through different ionization channels. We simply refer to the properties of the cation in the unrelaxed geometry of the neutral molecule.

Equations \eqref{eq:2} and
\eqref{eq:3} show very explicitly that the ionization potential
depends not only on the magnitude of the electric field vector
$\mathbf{F}$ but also on the angles of the field orientation with
respect to the principal polarizability axes and the permanent dipole moment
of the molecule. This is the essential ingredient that enters into our modification of the tunneling theory. Note that static Stark shifts were considered earlier in discussion of dissociation \cite{dietrich:1992:jcp,H.Akagi09112009}

The modification of the ionization potential due to Stark shifts results in the tunneling exponential $\exp (- 2 {\kappa}^3 ( \mathbf{F}) / 3F ) $, where the factor
\begin{equation}
\kappa ( \mathbf{F} ) = \sqrt{ 2 I_p ( \mathbf{F} ) }
\label{eq:kappa}
\end{equation}
dependens on the angles of field orientation with respect to the molecular axis. One can now in principle take the tunneling rate of Eq. \eqref{eq:4} and replace $\kappa$ with $\kappa (\mathbf{F} )$ everywhere. However by doing so, one has assumed that the initial orbital is not affected by the polarization of the molecule. This is in general not true, especially for field strengths at which the polarizability term gives much larger contribution to the Stark-shifted ionization potential of Eq. \eqref{eq:2} than the permanent dipole moment term. In these cases, the polarizability can modify the initial molecular orbital so that the $C_{lm}$ coefficients in Eq. \eqref{eq:5} become a function of the field strength. It is in general very hard to account analytically for such strong modifications of the initial orbital. However, as we show below by direct comparison to the experiment, in the case of OCS and in the limit of large fields it suffices to simplify the situation by disregarding the modifications of the HOMO and assuming that the angle-dependence of the tunnel emission occurs only due to the action of the Stark shift. Namely, we assume that the inner structure of the orbital is modified so much by the polarization response so that its asymptotic properties would be equivalent with respect to the axis defined by $\mathbf{F}$ for each orientation of the field with respect to the molecular axis. In other words, if the Stark shift would not be present, there would be no orientation dependent emission, i.e., the tunneling probability would be equal for all field orientations with respect to the molecular axis. Having this in mind, we can model this behaviour by taking an \textit{atomic} s-like state as an initial state in the tunneling model. This radical model will generally be better for large intensities, for systems with large polarizabilities and in cases where the initial orbital is not such that the polarization plane of the circularly polarized laser field lies entirely in the nodal plane of the orbital \cite{BN_circ}.

The  intensity of the laser pulse used in the experiment is very large so that for atoms with the same binding energy the ionization would occur over the barrier. To examine whether that holds for the case of the OCS molecule one requires a single-active-electron potential corresponding to the HOMO of OCS. Following the approach given in Refs. \cite{Abu-samhaPRA09,Abu-samhaPRA2010}, applied sucessfully to  CO$_2$ and other linear molecules, we have built a single-active-electron potential for OCS.

\begin{figure}
\includegraphics[width=\figwidth]{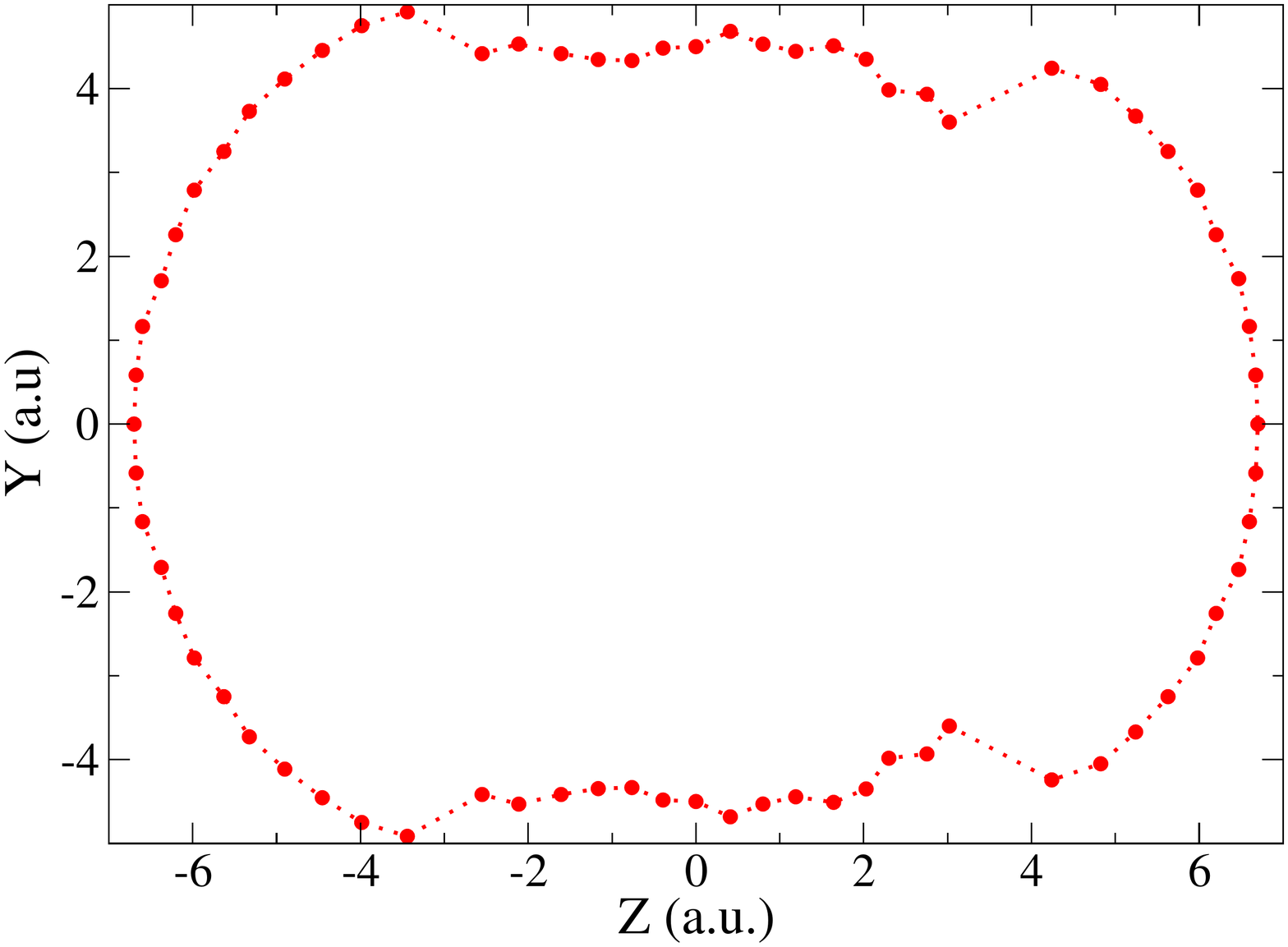}
\caption{(Color online) The saddle points of the OCS potential at the intensity of 2.44$\times 10^{14}$ W/${\rm cm^2}$ in the polarization ($Y$-$Z$) plane. The dots are the saddle points of the effective potential made up of the molecular single-active electron potential and the external field. The values of the saddle points are taken at 5 degree steps with respect to the z-axis, and are connected with the dotted lines to guide the eyes. The origin is the center of mass of the molecule. The S-end of the molecule is to the left of this figure, in accordance with the geometry of Fig. \ref{fig:PAD:schematic}. The explicit nuclear positions are as follows. S (-2 a.u.), C (0.99 a.u.), and O (3.2 a.u.) (see Appendix A).}
\label{fig:potential}
\end{figure}

Using the single-active-electron potential for OCS, we have verified that at the peak intensities of the experiment, the ionization occurs over the barrier at all angles of orientation of the field with respect to the molecular axis and both with or without inclusion of the Stark shift. The saddle points of the potential at the experimental peak intensity of 2.44$\times 10^{14}$ W/${\rm cm^2}$  are shown in Fig. \ref{fig:potential}. It can be seen that the saddle points occur relatively far away from the center-of-mass coordinate so the influence of the potential to the outgoing electron, born at the saddle points is very small. Hence the asymmetry of the molecular potential does not play a role in the present case.
In addition, due to the large polarizability of the parent ion (see Table \ref{tab2} in Appendix A), the induced dipole of the cation is very large and the orientation is such that it shields the electron from the influence of the attractive Coulomb potential, thereby decreasing the effect of the long-range Coulomb potential.

At over-the-barrier intensities, the tunneling rate given by the tunneling theory  overestimates the probability of ionization \cite{Tong:2005:JPB,Dimitrovski2008}. The saturation of the tunneling rate can be included by calculation of the exact, complex eigenenergies in a static field \cite{Dimitrovski2008,Solovev2000}, however this approach is limited only to very simple systems. In this study, we will adopt an ad-hoc and simple exponential factor, given in Ref. \cite{Tong:2005:JPB}, to account for the over-the-barrier saturation of the tunneling rate. The saturation factor due to over-the-barrier emission reads,
\begin{equation}
W ( \mathbf{F} )= w ( \mathbf{F} ) \exp \left( -6 \left(\frac{2}{{\kappa}^2 (\mathbf{F})} \right) \left( \frac{F}{{\kappa}^3 (\mathbf{F} ) } \right) \right) ,
\label{eq:7}
\end{equation}
where we have additionally included the Stark shifts in $\kappa$ according to Eq. \eqref{eq:kappa} since it occurs as an argument in the exponential. In the above equation, $w ( \mathbf{F} )$ is the tunneling rate from Eq. \eqref{eq:4}, where $C_{l,m}= \delta_{l,0} \delta_{m,0} $ corresponds to an $s$-state and $\kappa$ has been replaced by $\kappa ( \mathbf{F} )$. In summary, in the tunneling model we use to describe the momentum distributions for the OCS molecule, the tunneling rate is calculated as

\begin{align}
W (\mathbf{F}) = & \frac{1}{2 {\kappa ( \mathbf{F} )}^{\frac{2}{\kappa ( \mathbf{F} )}-1 }} \left( \frac{2 {\kappa ( \mathbf{F} ) }^3 }{F} \right)^{\frac{2}{\kappa (\mathbf{F} )} -1} \exp \left( -\frac{2 {\kappa ( \mathbf{F} )}^3 }{3 F} \right) \notag \\ 
& \times  \exp \left( -6 \left(\frac{2}{{\kappa ( \mathbf{F} ) }^2} \right) \left( \frac{F}{{\kappa (\mathbf{F} ) }^3} \right) \right) ,
\label{eq:8}
\end{align}
with $\kappa ( \mathbf{F} ) $  given by Eq. \eqref{eq:kappa}.

\section{Comparison with the experiment of strong-field ionization of 1D-aligned OCS}
\label{comparison}
We consider the experiment on strong-field ionization of 1D oriented OCS by a circularly polarized field in the $(Y,Z)$ plane. The molecular axis is along the $Z$-axis, and it is perpendicular to a detector that lies in the $(X,Y)$ plane (see Fig. \ref{fig:PAD:schematic}).

We assume the electric field of a LCP pulse as given by Eq.~\eqref{eq:9} and include the static Stark shift of the active HOMO through the shifts of the molecule and the unrelaxed cation as given in Eqs.~\eqref{eq:2}-\eqref{eq:3}. This approach leads to a modification of the ionization potential as a function of the angle of the direction of the field $\mathbf{F}$ with respect to the $Z$-axis (denoted as $ \theta = \mathbf{ \hat{F}}= \omega t$) as
\begin{align}
I_p( \theta )  =  & \; I_p (0) + ( {\mu}^{I} - {\mu}^{M} ) F_0 \cos ( \theta ) \notag \\ & +  \frac{1}{2} F_0^2  \left( ({\alpha}_{ZZ}^{M} - {\alpha}_{ZZ}^{I}) - ({\alpha}_{XX}^{M} - {\alpha}_{XX}^{I}) \right) {\cos}^2 (\theta ) \notag \\ & +   \frac{1}{2} F_0^2 ({\alpha}_{XX}^{M} - {\alpha}_{XX}^{I}).
\label{eq:11}
\end{align}
Note that for the angle $\theta$ of the field, the emission occurs in the direction $\theta + \pi$. The above equation gives the modification of the ionization potential as a function of the angle between the instantaneous direction of the field and the permanent dipole of the molecule, and together with Eq.~\eqref{eq:8} it
is the main theoretical input in the interpretation of the experiment.
The modification of the ionization potential is such that when the field vector and the permanent dipole moment of the molecule are parallel, the ionization potential is minimal so the tunneling probability reaches maximum. In that case the emission occurs opposite to the field direction, that is, from the O-end of the molecule. On the other hand, when the instantaneous field points from the S- to the O-end of the molecule, i.e., the field and the permanent dipole moment are antiparallel, the ionization potential is maximal and, conversely, the ionization probability minimal. The emission then occurs from the S-end of the molecule.

In the simplest model, the final momenta in the ($Y$-$Z$) plane are given by Eq. \eqref{eq:12}. Since there is larger probability of tunneling from the O-end of the molecule, for the LCP, there would be a larger probability that electrons with $p_Y > 0 $ appear, and for RCP pulse the situation would be opposite, which is in accord with the experiment. Scanning through instants of time within one cycle of the field, all possible final momenta $p_Y$ and $p_Z$ in the plane $p_X=0$ are reached and the associated ionization probability calculated from Eq. \eqref{eq:8}, where $| \mathbf{F} | = F_0$ and the orientation of $\mathbf{\hat{F}}$ is given by $\theta$ defined in Eq. \eqref{eq:12}. On the other hand, the transverse, $p_X$ component of the momentum cannot be changed by the external electromagnetic field and it is obtained from the well-known expression \cite{DeloneKrainov} for the momentum distribution of the transverse momenta of the tunneled electron at its birth, i.e.,
\begin{equation}
W ( p_X ) \sim \exp \left( - \frac{ \sqrt{2 I_p (0)}}{F_0} p^2_X \right).
\label{eq:13}
\end{equation}
Note that in actual calculations for the transverse distribution the approximation $I_p( \theta ) \approx I_{p} (0)$
has been used since the inclusion of Stark shifts in this degree of freedom does not alter the results significantly and has no influence on
the observed asymmetry in the experiment.

\begin{figure}
\begin{center}
\includegraphics[width=\figwidth]{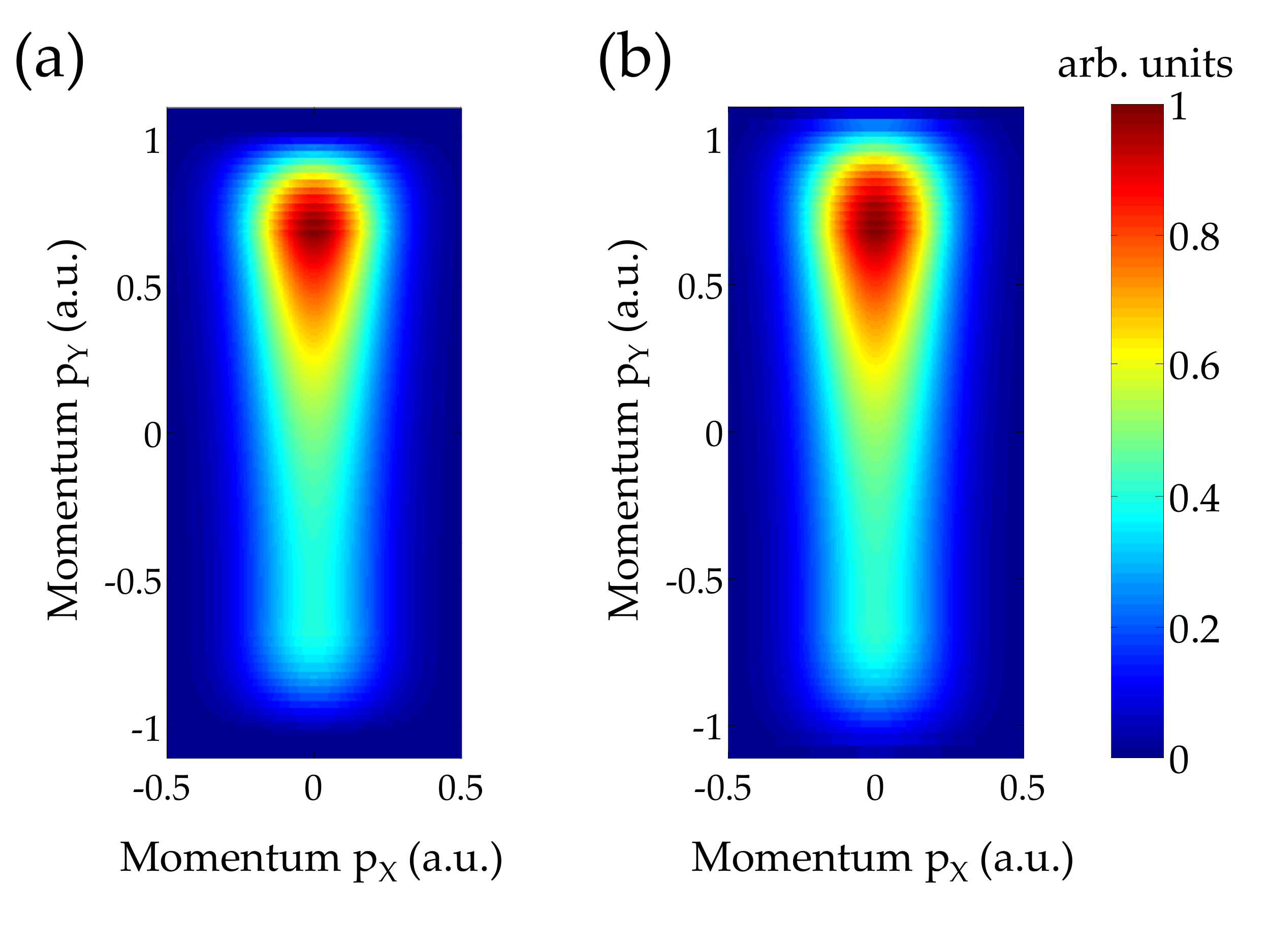}
\caption{(Color online) Momentum ditributions obtained by the present model for peak intensities of (a) $2.44 \times 10^{14}$ and (b) $ 2.83 \times 10^{14} $ W/${\rm cm}^2$ of the 800 nm, 30 fs probe laser pulse. Compare with the experiments in Figs. \ref{fig:OCS:PADs:intensities} (e) and (h).
} \label{fig:1}
\end{center}
\end{figure}

In the experiment, the measured quantity on the detector is
\begin{align}
W( p_X ,p_Y )= \int_{-\infty}^{\infty} d p_Z W( \mathbf{p} ) , \notag \\
\text{where} \quad W (\mathbf{p} ) = \frac{ {\partial}^3  P_{ion} }{ \partial p_X \partial p_Y \partial p_Z } ,
\label{eq:14}
\end{align}
and $P_{ion}$ is the total ionization probability. In the simpleman model, by inserting Eq. \eqref{eq:11} into \eqref{eq:8}, the tunneling rate $W ( \theta ) $ can be obtained, and from there, $W(0, p_Y ,  p_Z ) = W( \theta )$, where $p_Y$ and $p_Z$ dependency on $\theta$ is taken from \eqref{eq:12}. This momentum distribution is then integrated over $p_Z$ to obtain $w(0, p_Y )$. Having in mind Eq. \eqref{eq:13}, the momentum distribution \eqref{eq:14} is obtained as

\begin{equation}
W (p_X , p_Y ) = W(0, p_Y ) \exp \left( - \frac{ \sqrt{2 I_p (0)}}{F_0} p^2_X \right) .
\label{eq:15}
\end{equation}

As discussed in section \ref{alignment-orientation}, the orientation of the OCS molecules in the experiment is not perfect. In fact, 80\% of the molecules are oriented in the desired orientation (O-end towards detector) and 20\% oppositely. These 20\% of the molecules actively participate in the formation of the experimentally obtained momentum distribution hence this is an effect which must be taken into account in the theoretical model. The corresponding momentum distribution with this effect taken into account is readily obtained from \eqref{eq:15} as

\begin{equation}
W_{pn} (p_X , p_Y ) = 0.8 W ( p_X , p_Y ) + 0.2 W (p_X , - p_Y).
\label{eq:16}
\end{equation}

The final effect which must be taken into account to reproduce the experiment is the volume effect for tightly-focused laser beams. The consideration of this effect requires the calculation of momentum distributions of type \eqref{eq:16} for different intensities and then using a volume function to weight the momentum distributions at a particular intensity \cite{Wang:2005:OPL}. After inclusion of the volume effect, we obtain the momentum distributions given in Fig. \ref{fig:1} for the experimental peak intensities of $2.44 \times 10^{14}$ [\ref{fig:1}(a)] and $ 2.83 \times 10^{14} $ W/${\rm cm}^2$ [\ref{fig:1}(b)]. These momentum distributions visually resemble very much the corresponding experimental momentum distributions of Figs. \ref{fig:OCS:PADs:intensities} (e) and (h). As with  their experimental counterparts, the calculated momentum distributions in Figs. \ref{fig:1} (a) and (b) are very similar, reflecting that the momentum distributions are only weakly dependent on the laser intensity.

\begin{figure}
\begin{center}
\includegraphics[width=\figwidth]{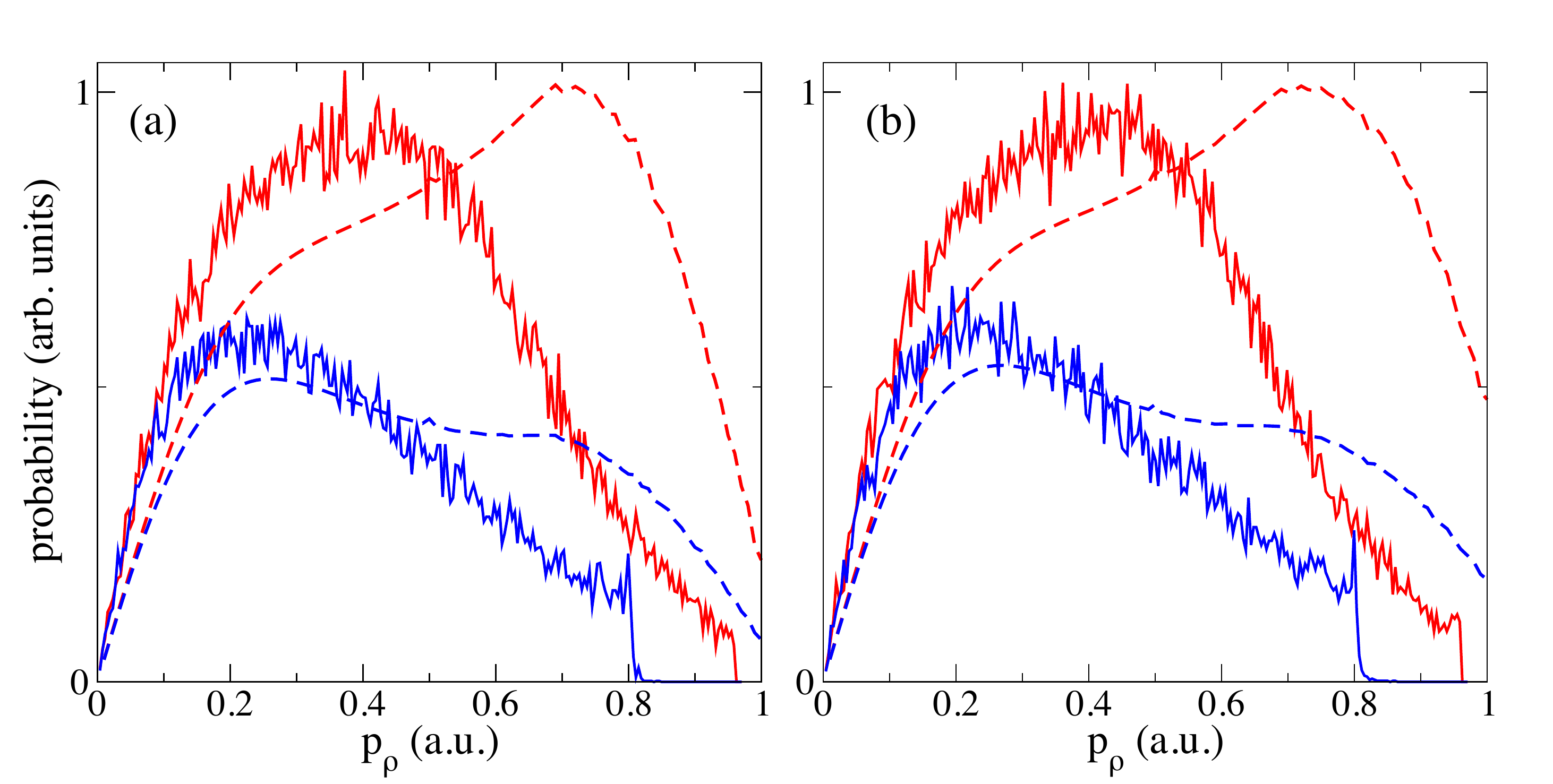}
\caption{(Color online) Comparison of experimental (full lines) and theoretical (dashed lines) radial distributions for peak intensities of (a) $2.44 \times 10^{14}$ and (b) $ 2.83 \times 10^{14} $ W/${\rm cm}^2$ of the 800nm, 30 fs probe laser pulse.  The upper curves are the radial distributions in the upper ($p_Y > 0$) and the lower curves are radial distribution in the $p_Y < 0$ half plane, see Eqs.~\eqref{eq:19}-\eqref{eq:20}.} \label{fig:2}
\end{center}
\end{figure}

A more detailed comparison between the experimental  [Figs \ref{fig:OCS:PADs:intensities} (e) and (h)] and theoretical momentum distributions [Figs. \ref{fig:1} (a) and (b)], reveals larger $Y$-components of the final momenta in the theoretical momentum distributions, see also Fig \ref{fig:2}. This is due to two reasons, which are not included in the theory presented here. First, the target OCS molecules are not ideally aligned with $\langle\cos^2\theta\rangle$~$\sim$~0.9. Hence, there is a non-vansihing probability that the angle between the molecular axis and the space-fixed $Z$-axis is nonzero. If we recall the post-ionization dynamics discussed in the previous section, the continuum electrons with the largest $Y$-component of the final momentum escape into the continuum at times when the electric field vector $\mathbf{F} (t)$ is parallel or antiparallel with the $Z$-axis [see Eq. \eqref{eq:12}]. On the other hand, emission from the O-end of the molecule is preferred and due to the nonzero value of the angle between the space-fixed $Z$-axis and the molecular axis, the absolute value of the $Y$-components of the final momenta would be smaller than the maximal value, i.e. $| p_Y | < F_0 / \omega $, so the $Y$-components of the momenta are lowered. The second reason for small $|p_Y |$ in the experimental momentum distributions is the interaction of the outgoing electron with its parent ion, which is beyond the simpleman model. Although in the case of a circularly-polarized laser field such post-ionization interaction is limited, the long-range Coulomb part of the potential  will act to decrease the magnitude of the final momenta due to the attractive forces at times immediately after ionization.

\begin{figure}
\begin{center}
\includegraphics[width=\figwidth]{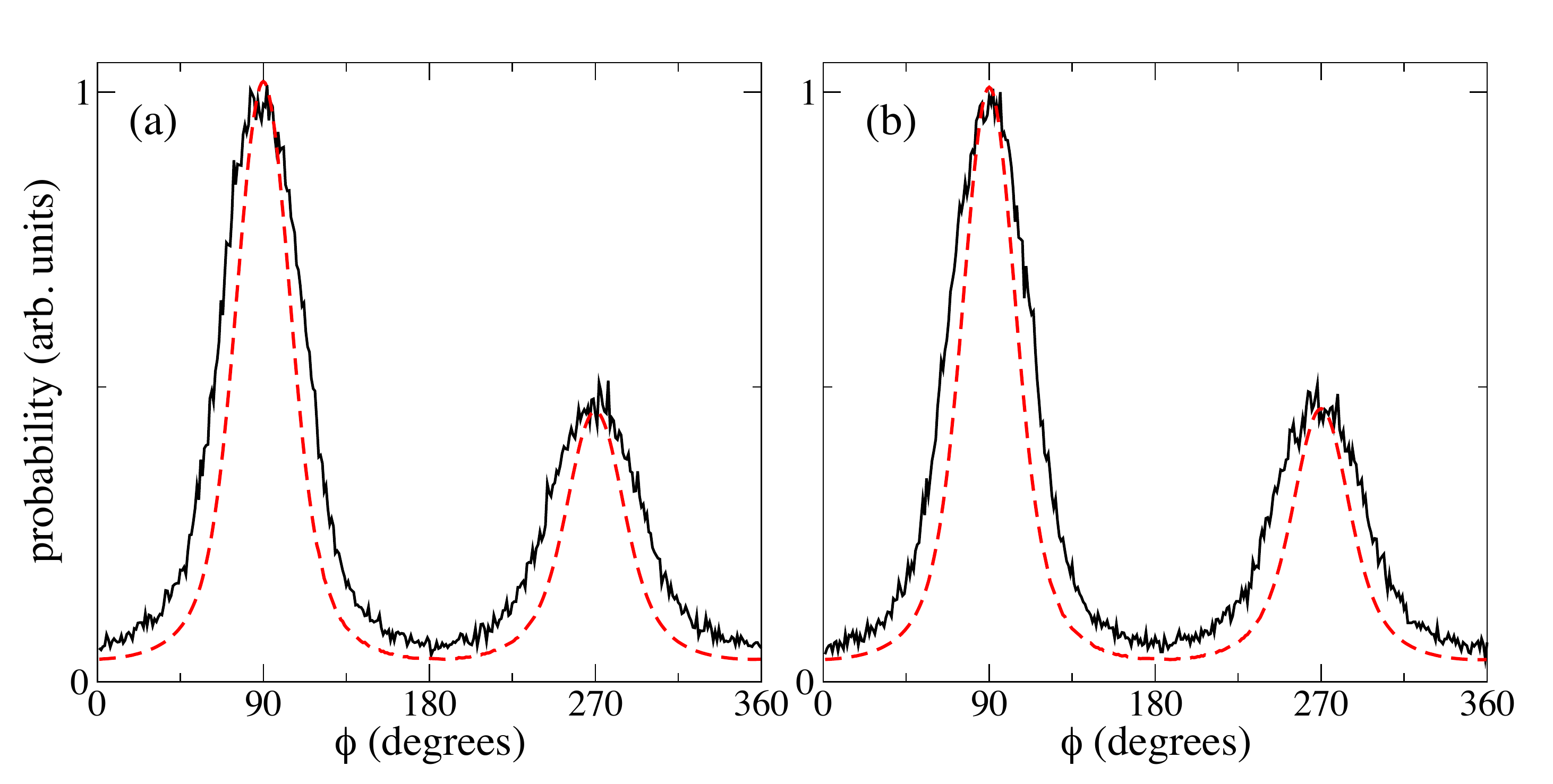}
\caption{(Color online) Comparison of experimental (full lines) and theoretical (dashed lines) angular distributions [Eq. \eqref{eq:18}] for peak intensities of (a) $2.44 \times 10^{14}$ and (b) $ 2.83 \times 10^{14} $ W/${\rm cm}^2$ of the 800 nm, 30 fs  probe laser pulse.} \label{fig:3}
\end{center}
\end{figure}

The above effects, however, have no influence on the most prominent feature observed in this experiment - the up-down asymmetry in the momentum distributions. This asymmetry  perpendicular to the permanent dipole moment is entirely described by including the static Stark shifts in the tunneling model. The up/total asymmetry $A^+ $ is defined as
\begin{eqnarray}
A^+ & = & P_{ion}^{Y>0}/P_{ion} \quad {\rm where} \nonumber \\
P_{ion}^{Y>0} & = & {\int}_{-\infty}^{\infty} dp_X {\int}_{0}^{\infty} dp_Y W_{pn} (p_X , p_Y) \quad {\rm and} \\
P_{ion} & = & {\int}_{-\infty}^{\infty} dp_X {\int}_{-\infty}^{\infty} dp_Y W_{pn} (p_X , p_Y) . \nonumber
\label{eq:17}
\end{eqnarray}
We have calculated $A^+ = 0.651$ (experimental value 0.64) at the intensity of $2.44 \times 10^{14}$ W/${\rm cm}^2$ and $A^+ = 0.649$ (experimental value 0.64) at the intensity of $2.83 \times 10^{14}$ W/${\rm cm}^2$. We conclude that the theoretical and the experimental asymmetry agree well.

\begin{figure}
\begin{center}
\includegraphics[width=\figwidth]{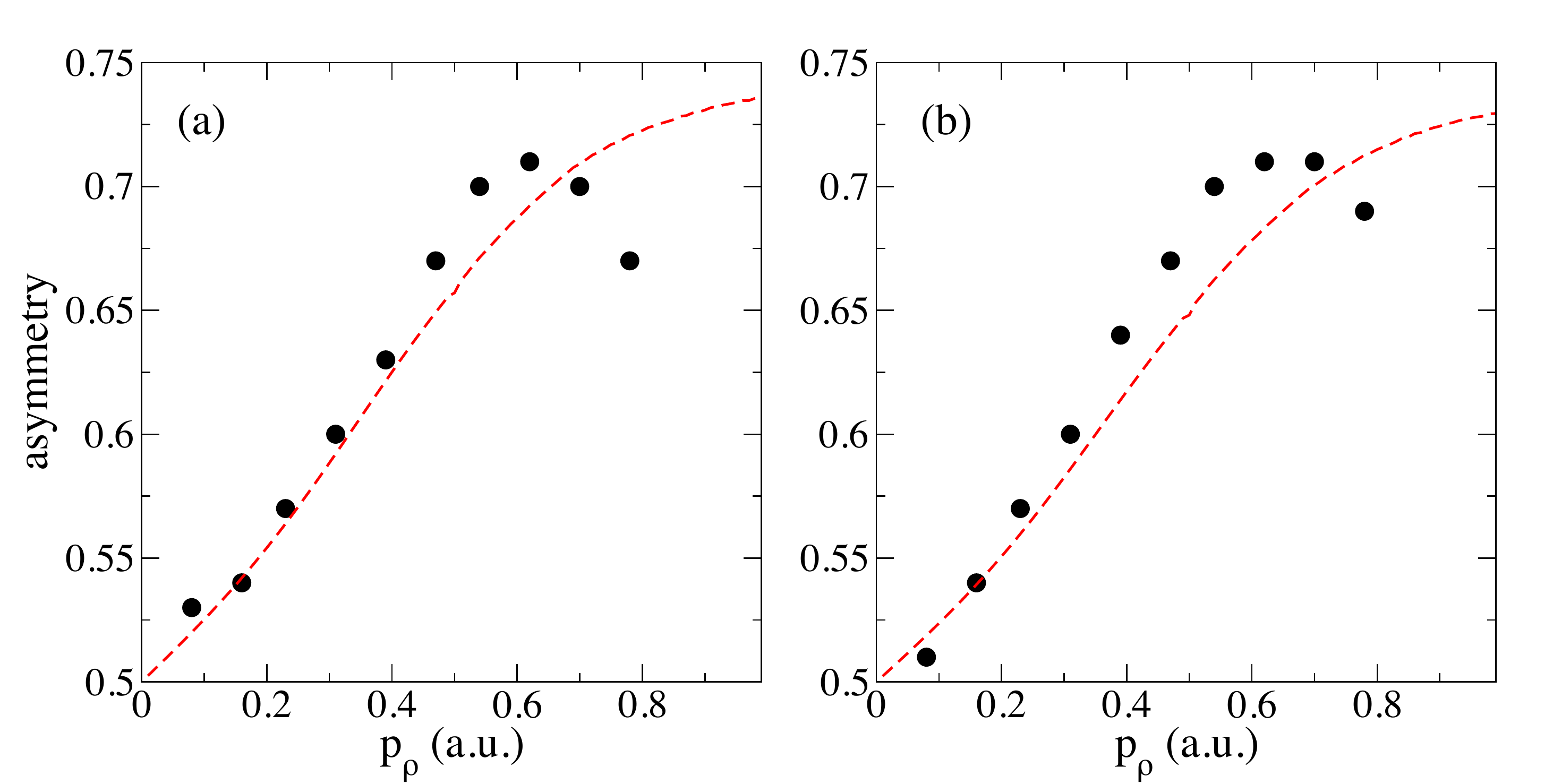}
\caption{(Color online) Comparison of experimental (dots) and theoretical (dashed lines) asymmetry $A^+$ [Eq. \eqref{eq:21}] as a function of radial momentum for peak intensity of (a) $2.44 \times 10^{14}$ and (b) $ 2.83 \times 10^{14} $ $ \text{W}/ \text{cm}^2$ of the 800 nm, 30 fs probe laser pulse.} \label{fig:4}
\end{center}
\end{figure}

We turn to the comparison of the differential quantities that can be derived from the momentum ditributions. Transforming the momentum distribution of Eq.~\eqref{eq:16} into polar coordinates $p_X=p_{\rho} \cos( \phi )$ and $p_Y = p_{\rho} \sin( \phi )$, one obtains the angular distribution
\begin{equation}
W( \phi ) = \int_0^{\infty} d p_{\rho} p_{\rho} W ( p_{\rho} , \phi )
\label{eq:18}
\end{equation}
shown in Fig. \ref{fig:3} and the radial distribution in the upper half ($p_Y > 0$)
\begin{equation}
W^{p_Y > 0} ( p_{\rho} ) = p_{\rho} {\int}_0^{\pi} d \phi W ( p_{\rho} , \phi )
\label{eq:19}
\end{equation}
and in the lower half of the plane ($p_Y < 0$)
\begin{equation}
W^{p_Y < 0} ( p_{\rho} ) = p_{\rho} {\int}_{\pi}^{2 \pi} d \phi W ( p_{\rho} , \phi )
\label{eq:20}
\end{equation}
shown in Fig. \ref{fig:2}.

The results for the angular distribution in Fig. \ref{fig:3} are in excellent agreement with the experiment. In particular the peak ratio of the peaks in the angular distribution for 90 and 270 degrees is reproduced by the theory. The width of the two peaks in the experimental radial distributions is slightly larger than the theoretical peaks. This is due to the non-perfect alignment of the molecule along the $Z$-axis discussed above. Moreover, one has to consider Coulomb focusing: the electron wavepacket is attracted to the $(Y,Z)$ plane, which results in a narrower width of the angular distributions. Thus the nonperfect alignment and the effect of the Coulomb focusing have an opposite effect on the angular distributions, almost cancelling each other and resulting in an overall very good agreement between theory and  experiment. On the other hand, as discussed above, the non-perfect alignment and the long-range interaction on the outgoing electron both have a tendency to decrease the final momenta. This is evident from the comparison of the experimental and theoretical radial distributions presented in Fig. \ref{fig:3}. The theoretical model overestimates the $Y$-components of final momenta. We note that the sudden cut-off of the experimental radial distributions is due to the maximum momentum that can be recorded on the detector, and not because of some physical effect.

The final quantity that has been calculated from the experimental data is the up/total asymmetry as a function of the radial momentum $A^+ ( p_{\rho} )$, defined as
\begin{equation}
A^+ ( p_{\rho} ) = \int_{0}^{ \pi } d \phi W ( p_{\rho}, \phi )/ \int_{0}^{ 2 \pi } d \phi W ( p_{\rho}, \phi ).
\label{eq:21}
\end{equation}
Experimental and theoretical results for this quantity  is presented in Fig.~\ref{fig:4}. For both intensities, the results for the differential asymmetry $A^+ ( p_{\rho} )$ agree very well in the region of smaller radial momenta. Namely, the differential asymmetry rises gradually from the value of 0.5 (no asymmetry) to around 0.7 at the peak. Very small radial momenta correspond to ionization when the electric field vector points in the direction of  the positive or negative $Y$-axis (electric field vector  perpendicular to the dipole moment) with equal probability of ionization  in both directions [put $\theta = \pi / 2 $ in Eq. \eqref{eq:11}], therefore no asymmetry. As the radial momentum $p_{\rho}$ increases, $|p_Y |$ increases and $ | \cos ( \theta) | $ of Eq. \eqref{eq:11} also increases, resulting in larger asymmetry. The theoretical model captures these features in the experimental data. The small discrepancy of 
experimental and theoretical results for $A^+ ( p_{\rho} ) $ at small and intermediate $p_{\rho}$ values is again due to the overestimation of the Y components of the final momenta by the theory, discussed above. At the largest $p_{\rho}$ values the decrease of the asymmetry in the experimental curves is not reproduced by 
theory. This could be due to nonperfect alignment or to focal 
volume effects, i.e., the asymmetry at 0.6 comes not only from 
the contribution of the tunneling slightly off the O end of the 
molecule at the peak intensity but also from tunneling exactly 
from the O end for lower peak intensities contributing to the 
total signal. Finally, the discrepancy could arise because the 
model neglects any orientation-dependent tunneling that has 
its origin in the initial state. 
\section{Conclusions}
We have studied photoionization of aligned and oriented OCS molecules. The prepared gas sample
was ionized by intense near-infrared femtosecond laser pulses. The detection of ionic fragments was used to investigate the degree of alignment and orientation. The sample was then used to investigate molecular frame photoelectron angular distributions (MFPADs) in the tunneling and over-the-barrier regimes.  Strong asymmetries in the distributions were observed and explained in terms of a modified tunneling theory. The circularly polarized field steers the electron away, and minimizes rescattering, but the presence of a permanent dipole moment and a large polarizability of the active orbital, the HOMO in the present case, means that the effective ionization potential shifts depending on the instantaneous magnitude and direction of the external field with respect to  the molecular axis.

In OCS and for the present set of laser parameters, the electron enters the continuum at such large distances that the asymmetry associated with the ionization potential in the initial tunneling process is sufficient to explain the experimental findings. The asymmetry associated with the molecular potential does not play a role, since the leading  asymmetric dipole term is suppressed at the distances in question [see Fig. 6].

It is our goal to extend the present techniques to study time-resolved electron dynamics, for example, to monitor changes in the MFPADs during a photochemical reaction. In such a process the nuclear motion could be sufficiently slow  that changes in the MFPADs will occur at the femtosecond timescale~\cite{bisgaard:Science:2009}, and changes could then be recorded by firing a short femtosecond pulse. In the tunneling regime, asymmetries in the MFPAD will be exponentially sensitive to the ionization potential and hence to  changes in the dipole moments and polarizabilities. Changes in the MFPADs will consequently link directly to the instantaneous values of these quantities. There might be changes in the MFPADs for reasons other than the ones connected with the dipoles and polarizabilities. We mentioned asymmetries due to the exact form of the molecular potential above. Another possibility is the temporal formation and changes of nodal surfaces. It is encouraging that  signatures of nodal surfaces is clearly detectable with the present technique, as shown for benzonitrile~\cite{Holmegaard:NatPhys:2010,BN_circ}, and as discussed theoretically in Ref.~\cite{Martiny2010}.

\section{Acknowledgement}
\label{acknowledgement}

We thank C.P.J. Martiny for useful discussions.
The work was supported by the Danish National Research Foundation, the Lundbeck Foundation, The Carlsberg Foundation, the Danish Council for Independent Research (Natural Sciences) and the European Marie Curie Initial Training Network Grant No. CA-ITN-214962-FASTQUAST.

\section*{APPENDIX A: Molecular properties of OCS}

The molecular properties of OCS were obtained for the experimental
geometry~\cite{OCS-exp} based on the Hartree-Fock
wavefunction~\cite{GAMESS}, with the molecule oriented along the
$Z$-axis and the O-end pointing towards the detector
(see Fig. \ref{fig:PAD:schematic}). In the coordinate system, defined in Fig. \ref{fig:PAD:schematic}, where the center of mass of OCS is the origin, the $Z$-coordinates of the atomic centers are -2 a.u. for S, 0.99 a.u. for C, and 3.2 a.u. for the O atom. The first ionization potential
($I_p$), computed as the positive energy of the highest occupied
molecular orbital (HOMO), is 11.4~eV (0.42 a.u.), in agreement
with the experimental value of 11.2~eV~\cite{OCS-exp-ip}. The OCS
molecule has an asymmetric charge distribution with a dipole
moment of 0.71~D (1~a.u.=2.54~D) pointing towards the
S-end \cite{tanaka:2835}. The OCS molecule has two degenerate HOMO orbitals shown in
Fig.~\ref{fig:OCS_orbitals}. By expanding the total wave function in a linear combination of atomic orbitals $\phi_i$, centered at each nucleus $\Psi=\sum_i c_i \phi_i$, we obtain an estimate of the electron population at each center as $(c_i)^2$. This population analysis indicates that 75\% of
the HOMO electron density is localized on the S-end, 15\% on the
O-end and only 10\% on the C atom. The orbital lying just below the HOMO in energy has a significantly higher ionization potential (17.1~eV
from Hartree-Fock calculations \cite{GAMESS}), and since we are in the tunneling regime with an exponential sensitivity to the ionization potential its contribution to the
ionization dynamics is expected to be negligible.

\begin{table}

\caption{The dipole moment of the OCS molecule and the cation (in
the geometry of the neutral molecule and both pointing from the
O-end to the S-end), given in units of Debye (1~au=2.54~Debye).}

\begin{tabular}{lcccccc}
\hline
  \multicolumn{1}{c}{State}&
  \multicolumn{1}{c}{TZV\footnote{Valence triple-$\zeta$ basis set~\cite{jr.:716}.}}&
  \multicolumn{1}{c}{ACCT\footnote{Dunning triple-$\zeta$ basis set~\cite{jr.:1007}.}}&
  \multicolumn{1}{c}{PC\footnote{Jensen polarization-consistent basis set~\cite{jensen:9113}.}} &
  \multicolumn{1}{c}{APC\footnote{Jensen polarization-consistent basis set with diffuse basis functions~\cite{jensen:9234}.}} &
  \multicolumn{1}{c}{Sadlej\footnote{Sadlej pVTZ basis set~\cite{Sadlej}.}} \\
        \hline
OCS     &  0.52  & 0.72 & 0.71 & 0.88 & 0.65 \\
OCS$^+$ &  1.83  & 2.69 & 2.00 & 2.20 & 1.61 \\
    \hline
\end{tabular}

\label{tab1}
\end{table}

\begin{table}

\caption{The non-zero components of the polarizability and
hyperpolarizability for the OCS molecule and the cation (in the
geometry of the neutral molecule), given in atomic
 units.}

\begin{tabular}{lrrrrrrrrrrrr}
\hline
  \multicolumn{1}{c}{State}&
  \multicolumn{2}{c}{$\alpha_{XX}$}&
  \multicolumn{2}{c}{$\alpha_{YY}$}&
  \multicolumn{2}{c}{$\alpha_{ZZ}$}&
  \multicolumn{2}{c}{$\beta_{XXZ}$}&
  \multicolumn{2}{c}{$\beta_{YYZ}$}&
  \multicolumn{2}{c}{$\beta_{ZZZ}$}   \\
        \hline
OCS     &  26.15  & & 26.15 & &  50.72  & & -45.92 & & -45.92 & & -12.85 & \\
OCS$^+$ &  19.06  & & 18.73 & &  44.09  & & -17.55 & & -20.17 & &  18.49  &\\
    \hline
\end{tabular}

\label{tab2}
\end{table}

In the development of the tunneling theory, we need the dipole
moments, polarizabilities and hyperpolarizabilities of the OCS
molecule. These were first computed using the Hartree-Fock
wavefunction in conjunction with the aug-cc-pVTZ~\cite{jr.:1007}
basis set. Comparisons to results reported for the neutral
molecule
 in the Computational Chemistry Comparison and Benchmark DataBase~\cite{CCDB}, suggest
that while the polarizabilities and hyperpolarizabilities are
reasonable, the dipole moments should be improved. We carried out
calculations using the MP2 level of theory with different basis
sets,
 and the resultant dipole moments are shown in Table~\ref{tab1}. For the neutral molecule,
the computed dipole moments are in fair agreement with
the experimental value (0.71~D~\cite{tanaka:2835}). For the
cation, on the other hand, we obtained a wider range of values
depending on the basis set. A simple classical model based on the
redistribution of atomic charges due to removal of the HOMO
electron, without allowing the remaining electrons to relax,
predicts $\mu_Z$~=~-2.2~D for the cation. This is in very good
agreement with the results based on the Jensen's
polarization-consistent basis sets (~-2.0~Debye without diffuse
functions~\cite{jensen:9113} vs. -2.2~Debye with diffuse
functions~\cite{jensen:9234}). In the tunneling model presented in
Section \ref{theory} we use the dipole moment for OCS$^+$ obtained
via the simple charge model. The molecular polarizabilities and hyperpolarizabilities
are shown in Table~\ref{tab2}.

\end{document}